\documentstyle[sprocl,psfig,epsfig]{article}
\def\be{\begin{equation}}
\def\ee{\end{equation}}
\def\bea{\begin{eqnarray}}
\def\eea{\end{eqnarray}}
\def\simlt{\stackrel{<}{{}_\sim}}
\def\simgt{\stackrel{>}{{}_\sim}}

\begin{document}
\begin{flushright}
SCIPP/97-19\\
IFT/97-13\\
hep--ph/9707497 \\
\end{flushright}
\title{SUPERSYMMETRIC LOOP EFFECTS \footnote{To appear 
in ``Perspectives on Supersymmetry'' 
G.L. Kane editor, World Scientific, Singapore 1997.}}
\author{PIOTR H. CHANKOWSKI \footnote{On leave of absence from the
Institute of Theoretical Physics, Warsaw University, Ho\.za 69, 
00-681 Warsaw, Poland.}}
\address{Santa Cruz Institute for Particle Physics\\
University of California, Santa Cruz, CA 95064, U.S.A.}
\author{STEFAN POKORSKI}
\address{Inst. of Theoretical Physics, Warsaw University, Ho\.za 69, 
00-681 Warsaw, Poland\\
and\\
Laboratoire de Physique Theorique, Univ. de Paris-Sud
91400, Orsay, France.}
\maketitle\abstracts{We review the loop effects of the low energy 
supersymmetry. The global success of the Standard Model rises two related
questions: how strongly the mass scales of the superpartners are constrained 
and can they be, nevertheless, indirectly seen in precision measurements. The 
bulk of the electroweak data is well screened from supersymmetric loop effects,
due to the structure of the theory, even with superpartners generically light,
${\cal O}(M_Z)$. The only exception are the left-handed squarks of the third 
generation which have to be $\simgt {\cal O}(300$ GeV) to maintain the success
of the SM. The other superpartners can still be light, at their present 
experimental mass limits, and would manifest themselves through virtual 
corrections to the small number of observables such as $R_b$, 
$b\rightarrow s\gamma$, $K^0$-$\bar K^0$ and  $B^0$-$\bar B^0$ mixing and a 
few more for large $\tan\beta$. Those effects require still higher experimental
precision to be detectable.}

\section{Introduction}
The masses of the weak bosons, $W^\pm$ and $Z^0$, as well as fermion masses 
originate from the mechanism of spontaneous symmetry breaking. This mechanism
requires the presence of elementary or composite scalar modes, the so-called
Higgs modes which develop nonvanishing vacuum expectation values. In the 
Standard Model (SM), viewed as an effective low energy theory, the Higgs 
potential looks very unnatural and the theory faces the well known hierarchy
problem. In brief, scalar potential is generically unstable with respect to 
quantum corrections from any new physics (the mass squared parameter of the 
potential receives loop corrections proportional to masses squared of the 
heavy new particles). Thus, the structure of the vacuum in the SM is strongly
suggestive of the existence of new scale of fundamental interactions, the
physical cut-off to the SM, close to the electroweak scale.

Supersymmetry offers an interesting solution to the hierarchy puzzle and,
moreover, has several other theoretical and phenomenological (gauge
coupling unification) virtues. The new scale,  the mentioned earlier cut-off
to the SM, is the scale of soft supersymmetry breaking. In other words, this 
is the scale (often it can be defined only in some average sense) of the mass 
spectrum of the superpartners to the particles of the SM. Two immediate and 
most important remarks about the superpartner spectrum are the following ones:
if supersymmetry is to cure the hierarchy problem that scale is 
expected to be not much above the electroweak scale. On the other hand, it is
totally unknown in detail, as we do not have at present any realistic model
of supersymmetry breaking. Therefore, the minimal 
supersymmetric extension of the SM, the so-called Minimal Supersymmetric
Standard Model (MSSM) is a very well defined theoretical framework but 
contains many free parameters: superpartner soft 
masses and their dimensionful
couplings. 

Lack of any detailed knowledge about the superpartner masses has obvious 
implications for the direct search for superpartners which can only be based
on systematic exploration of the higher and higher energy scales. It is,
therefore, very interesting to discuss the question to what extent the
superpartner spectrum can manifest itself through virtual (loop) effects 
on the electroweak observables. Do very high precision measurements of the 
electroweak observables  provide  us with a tool to see supersymmetric 
effects indirectly or, at least, to put stronger limits on its spectrum? 
We remember the important r\^ole played by precision measurements in seeing,
indirectly, some evidence for the top quark long ago its direct discovery and
with the mass quite close to its measured mass. Also, the present level of
precision makes the electroweak measurements to some extent sensitive even 
to the Higgs boson mass, although the dependence is only logarithmic. With
supersymmetric corrections the situation is different. The dependence on the 
top quark (and Higgs) mass in the SM is due to {\sl nondecoupling} of heavy 
particles which get their masses through the mechanism of spontaneous symmetry
breaking. The soft SUSY breaking is explicit and the Appelquist-Carazzone 
theorem \cite{APCA} applies to the superpartner spectrum. Thus, 
supersymmetric virtual effects dissapear at least as ${\cal O}(1/M_{SUSY})$. 
Nevertheless, several interesting questions can be discussed and this is the 
content of this Chapter.
First, we discuss the impact of the general succes of the SM in describing the
precision data on the existence of new physics and on supersymmetry in 
particular. The resulting constraints on the SUSY spectrum are reviewed 
with emphasize on the existence of the room for very light, ${\cal O}(M_Z)$, 
particles. Indeed, most of the superpartners effectively decouple from most 
of the electroweak observables much faster than ${\cal O}(1/M_{SUSY})$. This 
high degree of screening follows from the basic structure of the theory. 
There are only few exceptions to this general rule: effects of light, 
${\cal O}(M_Z)$, charginos, 
stops and the charged Higgs boson can be substantial in some specific 
observables like $R_b\equiv\Gamma(Z^0\to\bar bb)/\Gamma(Z^0\to hadrons)$ and
some flavour changing neutral current (FCNC) processes. In addition, for 
large $\tan\beta$ sizeable loop corrections to the Yukawa couplings from
those particles are also possible. Those effects are discussed in subsequent 
sections. The Chapter ends with a brief summary of the overall prospects.

\section{Supersymmetry and the electroweak precision data}

The bulk of the electroweak precision measurements ($M_W$, $Z^0$-pole 
observables, $\nu e$, $ep$ scattering data, etc.) shows that the global 
comparison of the SM predictions with the data is impressive. Both, the 
experiment and the theory have at present similar accuracy, typically 
${\cal O}(1$ $^{_{0}}\!\!\slash\!_{_{00}}$)! The predictions of the SM are 
usually given in terms of the very precisely known parameters $G_\mu$, 
$\alpha_{EM}$, $M_Z$ and the other three parameters $\alpha_s(M_Z)$, $m_t$, 
$M_h$. The top quark mass and the strong coupling constant are now also
reported from independent experiments with considerable precision: 
$m_t=(175.6\pm5.5)$ GeV and $\alpha_s(M_Z)=0.118\pm 0.003$, but those
measurements are difficult and it is safer to take $\alpha_s$, $m_t$, 
$M_h$ as parameters of an overall fit. Such fits give values of $m_t$ and 
$\alpha_s$ very well consistent with the above values 
\cite{ELLISY,ALBA,BLON,LEPEWWG97}.

The theoretical uncertainties in the SM predictions (for fixed $m_t$, $M_h$, 
$\alpha_s$) come mainly from the RG evolution of 
$\alpha_{EM}\equiv\alpha(0)\rightarrow\alpha(M_Z)$
which depends on the hadronic contribution to the photon vacuum polarization 
$\alpha(s)=\alpha (0)/(1-\Delta\alpha(s))$ where
$\Delta\alpha(s)=\Delta\alpha_{hadr}+\dots$ and
$\Delta\alpha_{hadr}=0.0280\pm 0.0007$ \cite{EID}. The present error in 
the hadronic vacuum polarization propagates as 
${\cal O}(1$ $^{_{0}}\!\!\slash\!_{_{00}}$) errors in the final predictions.
The other uncertainties come from the neglected higher order corrections and 
manifest themselves as renormalization scheme dependence, higher order 
arbitrariness in resummation formulae etc. Those effects have been estimated 
to be smaller than ${\cal O}(1$ $^{_{0}}\!\!\slash\!_{_{00}}$), hence the 
conclusion is that the theory and experiment agree with each other at the 
level of ${\cal O}(1$ $^{_{0}}\!\!\slash\!_{_{00}}$) accuracy. In particular, 
the genuine weak loop corrections are now tested at ${\cal O}(5\sigma)$ level 
and the precision is already high enough to see some sensitivity to the Higgs 
boson mass.

The electroweak observables depend only logaritmically on the Higgs boson 
mass (whereas the dependence on the top quark mass is quadratic). Global fits 
to the
present data give $M_h\approx130^{+130}_{-70}$ GeV and the 95$\%$ C.L. upper 
bound is around 470 GeV \cite{ELLIS,LEPEWWG97}. Thus, the data give some 
indication for a light Higgs boson. (It is worth noting that $M_h=1$ TeV is 
more than $\simgt3\sigma$ away from the best fit). The direct experimental 
lower limit on the SM Higgs boson mass $M_h$ is $\sim 70$ GeV. 

The overall global succes of the SM is a bit overshadowed by a couple of 
(quite relevant!) scattered clouds. The results for $R_b$ are still 
preliminary and differ by about 2$\sigma$ between different experiments. The 
world average is only 1.8$\sigma$ away from the SM prediction but with the 
error which is only slightly smaller than the maximal possible enhancement of 
$R_b$ in the MSSM (see later). The effective Weinberg angle is now reported 
with a very high precision. However, this result comes from averaging over 
the SLD and LEP results which are more then 3$\sigma$ apart. Hoping for 
further experimental clarification of those few points it is interesting to 
discuss already now the impact of the general succes of the SM on the 
existence of new physics. The simplest interpretation of the success of the 
SM within the MSSM is that the superpartners are heavy enough to decouple 
{}from the electroweak observables. Explicit calculations (with the same 
precision as in the SM) show 
that this happens if the common supersymmetry breaking scale is $\geq {\cal O}
(300-400)$ GeV \cite{ABC,ELFOLI,CP96.PL,DABFIT}. This is very important as 
such a scale of supersymmetry breaking is still low enough for supersymmetry 
to cure the hierarchy problem. However, in this case the only supersymmetric 
signature at the electroweak scale and just above it is the Higgs sector with 
a light, $M_h\leq {\cal O}(150)$ GeV, Higgs boson. This prediction is 
consistent with the SM fits discussed earlier \cite{ELLIS,LEPEWWG97}. 
We can, therefore, conclude at this point that 
the supersymmetric extension of the SM, with all superpartners 
$\geq {\cal O}(300)$ GeV, is phenomenologically as succesful as the SM itself 
and has the virtue of solving the hierarchy problem. Discovery of a light
Higgs boson is the crucial test for such an extension.

The relatively heavy superpartners discussed in the previous paragraph are 
sufficient for explaining the success of the SM. But is it necessary that all 
of them are that heavy? Is there a room for some light superpartners with 
masses ${\cal O}(M_Z)$ or even below? This question is of great importance 
for LEP2. Indeed, a closer look at the electroweak observables shows
that the answer to this question is positive. The dominant quantum corrections
to the electroweak observables are the so-called "oblique" corrections to the 
gauge boson self-energies. They are economically summarized in terms of the 
$S,T,U$ \cite{STU,ALBA} parameters 
\begin{eqnarray}
\alpha S\sim \Pi^\prime_{3Y}(0)=\Pi^\prime_{L3,R3} +\Pi^\prime_{L3,B-L}
\end{eqnarray}
(the last decomposition is labelled by the
$SU_L(2)\times SU_R(2)\times U_{B-L}(1)$ quantum numbers \cite{JAPP})
\begin{eqnarray}
\alpha T\equiv\Delta\rho\sim\Pi_{11}(0)-\Pi_{33}(0)
\end{eqnarray}
\begin{eqnarray}
\alpha U\sim \Pi^\prime_{11}(0)-\Pi^\prime_{33}(0)
\end{eqnarray}
where $\Pi_{ij}(0)$ $(\Pi^\prime_{ij}(0))$ are the (i,j) left-handed gauge 
boson self-energies at the zero momentum (their derivatives) and the 
self-energy correction to the $S$ parameter mixes $W^3_\mu$ and $B_\mu$ 
gauge bosons. It is clear from their definitions that the parameters $S,T,U$ 
have important symmetry properties: $T$ and $U$ vanish when 
quantum corrections to the left-handed gauge boson self-energies leave 
unbroken "custodial" $SU_V(2)$ symmetry. The parameter $S$ vanishes if 
$SU_L(2)$ remains an exact symmetry (notice that, since 
${\bf 3_L}\otimes {\bf 3_R}={\bf 1}\oplus {\bf 5}$ under $SU_V(2)$, exact 
$SU_V(2)$ is not sufficient for vanishing of $S$ \cite{JAPP}). 

In terms of the parameters $S$, $T$ and $U$ the ``new physics'' contribution
to the basic electroweak observables can be approximately written as 
\footnote{We assume that the supersymmetric contributions of order 
$M_Z^2/M^2_{SUSY}$ to those observables are small and can be neglected. 
If it is not the case, the parameters $S$, $T$ and $U$ should be defined at 
non-zero momentum transfer \cite{MYDR} and one should take into account also 
the ``new physics'' contributions through additional parameters like 
$\Delta\alpha=\Pi^\prime_{\gamma\gamma}(M_Z)-\Pi^\prime_\gamma(0)$ and 
$e_5=M_Z^2 F^{\prime}_{ZZ}(M_Z^2)$ \cite{BACAFR}.}
\begin{eqnarray}
\delta M_W = {M_W\over2}{\alpha\over c_W^2-s_W^2}\left(
c^2_W T^{new} - {1\over2}S^{new}
+{c^2_W-s^2_W\over4s^2_W}U^{new}\right)
\label{eqn:stu_mw}
\end{eqnarray}
\begin{eqnarray}
\delta\sin^2\theta^{eff}_{lept} =-{s^2_Wc^2_W\over c^2_W -s^2_W}\left(
\alpha T^{new}-{\alpha\over4s^2_Wc^2_W}S^{new}\right)
\label{eqn:stu_sin}
\end{eqnarray}
\begin{eqnarray}
\delta\Gamma(Z^0\rightarrow\overline ff)&=&{\alpha M_Z\over12s^2_Wc^2_W}
\left[\left(g^2_V + g^2_A\right)\left(
\alpha T^{new}\right)\right.\nonumber\\
&-&\left.4g_VQ^f{s^2_Wc^2_W\over c^2_W-s^2_W}\left(-
\alpha T^{new} - {\alpha\over4s^2_Wc^2_W}S^{new}\right)\right]
\label{eqn:stu_gam}
\end{eqnarray}
where the parameters $M_W$, $c_W\equiv M_W/M_Z$, $s_W$, 
$g_V=-I_3^f+2Q^f\sin^2\theta^{eff}_{lept}$, $g_A=-I_3^f$ are computed in the 
SM (taking into account loop corrections) and with some reference values of 
$m_t$ and $M_h$. $S^{new}$, $T^{new}$, $U^{new}$ contain only the 
contributions from physics beyond the SM. 

The success 
of the SM means that it has just the right amount of the $SU_V(2)$ breaking 
(and of the $SU_L(2)$ breaking), encoded mainly in the top quark-bottom quark 
mass splitting. Any extension of the SM, to be consistent with the precision 
data, should not introduce additional sources of large $SU_V(2)$ breaking in 
sectors which couple to the left-handed gauge bosons. In the MSSM, the main 
potential origin of new $SU_V(2)$ breaking effects in the left-handed sector 
is the splitting between the left-handed stop and sbottom masses
\cite{HABER,MYDR}:
\begin{eqnarray}
M^2_{\tilde t_L}&=&m^2_Q+m^2_t-{1\over6}\cos2\beta (M^2_Z-4M^2_W)\nonumber\\ 
M^2_{\tilde b_L}&=&m^2_Q+m^2_b-{1\over6}\cos2\beta (M^2_Z+2M^2_W)
\end{eqnarray}
The $SU_V(2)$ breaking is small if the common soft mass $m^2_Q$ is large 
enough. So, from the bulk of the precision data one gets a lower bound on the 
masses of the left-handed squarks of the third generation \footnote{Additional 
source of the $SU_V(2)$ breaking is also in the $A$-terms. In principle, there
can be cancellations between the soft mass term and the $A$-term contribution,
such that another solution with small $SU_V(2)$ breaking exists with a large 
inverse hierarchy $m^2_U\gg m^2_Q$. This is very unnatural from the point of 
view of the GUT boundary conditions and here we assume $m^2_Q>m^2_U$.}. 
However, the right-handed top and bottom squarks can be very light, at their 
experimental lower bounds $\sim 70$ and $\sim 150$ GeV, respectively. 

The other possible source of  $SU_V(2)$ violation is in the slepton sector.
For small values of $m^2_L$ the splitting between the left-handed slepton 
and sneutrino masses 
\begin{eqnarray}
M^2_{\tilde\nu}&=&m^2_L  +{1\over2}\cos2\beta M^2_Z\nonumber\\ 
M^2_{\tilde l} &=&m^2_L+m^2_l+{1\over2}\cos2\beta (M^2_Z-2M^2_W)
\label{eqn:leptsplit}
\end{eqnarray}
becomes non-negligible. Since this mass splitting may be of similar magnitude 
for all three generations of sleptons this effect should also be considered
\cite{MYDR}. It is also worth noting different behaviour of the slepton and
squark contributions to the $SU_V(2)$ breaking: the former vanishes in the 
limit $\tan\beta\rightarrow1$ whereas the latter is maximal in this limit and 
slightly decreases as $\tan\beta\rightarrow\infty$ \footnote{It is also worth 
mentioning that the $SU_V(2)$ breaking in the sector of the first two 
generation left-handed squarks is similar to that in the slepton sector 
(but enhanced by the colour factor of 3) i.e. it is determined by the 
``electroweak'' terms in their mass formulae (the mass squared splittings of 
the up- and down-type left-handed squarks are proportional to 
$\approx\cos2\beta M_W^2$). Such effects can be used to constrain from below
the soft SUSY breaking terms in the case of broken $R-$parity or if the gluino 
is heavier than ${\cal O}(350)$ GeV (i.e. when the direct Tevatron bounds, 
$M_{\tilde q}\simgt150$ GeV, for $\tilde q\neq\tilde t$, do not apply).}.

\begin{figure}
\psfig{figure=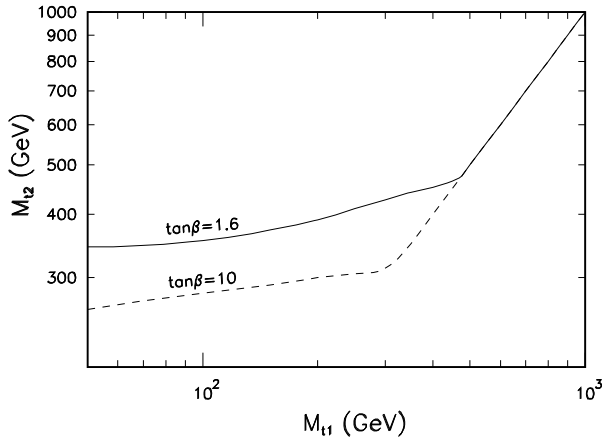,height=6.5cm}
\caption{Lower bounds on the heavier stop mass $M_{\tilde t_2}$, as a function
of $M_{\tilde t_1}$ for $\tan\beta=1.6$ (solid line) and
$\tan\beta=10$ (dashed line). A scan over the top quark mass and the 
top squarks mixing angle $\theta_{\tilde t}$ has been performed.}
\label{fig:kane1}
\end{figure}

In Fig. \ref{fig:kane1} we show the lower bound on the mass of the heavier 
top squark as a function of the mass of the lighter stop,  which follows from 
the requirement that a fit in the MSSM is at most by $\Delta\chi^2 =2$ worse 
than the analogous fit in the SM. 
{}From the analysis of the SUSY contributions to the parameters $T$ and $S$ it
follows that in the MSSM $T^{new}$ is always positive whereas $S^{new}$ is 
always negative \footnote{$S^{new}$ could be positive only in the small window
of the chargino parameter space which is already excluded by the unsuccesfull 
LEP search.}. Therefore, the full fits to the electroweak observables give 
more restrictive limits on the MSSM parameter space than do e.g. bounds on 
the $\Delta\rho(0)$ parameter alone because, as follows from eqs
(\ref{eqn:stu_mw}-\ref{eqn:stu_gam}), the effects in the $T$ and $S$ always 
add up. In the context of Fig. \ref{fig:kane1} there is one more interesting
observation to be made. In the low $\tan\beta$ region, for a given 
$M_{\tilde t_1}$, an absolute lower bound on $M_{\tilde t_2}$ is set by the 
(conservative) experimental lower bound on the lightest supersymmetric Higgs 
boson mass, $M_h\simgt 60$ GeV. For low $\tan\beta$, the tree level Higgs 
boson mass is close to zero and radiative corrections are very important. They 
depend logarithmically on the product $M_{\tilde t_1}M_{\tilde t_2}$. Also, 
since the best fit in the SM requires $M_h\approx130$ GeV, too small values 
of $M_{\tilde t_2}$ (leading to too small value of $M_h$) are disfavoured by
the MSSM fit. The limit shown in Fig. \ref{fig:kane1} take both effects into 
account. They explain the difference between the limits for small and large
$\tan\beta$ cases. The absolute limits on the stop masses obtained from the 
bound on $M_h$ (which apply only for low $\tan\beta$) turns out to be slightly
weaker than the limits from the fit shown in Fig. \ref{fig:kane1}.

\begin{figure}
\psfig{figure=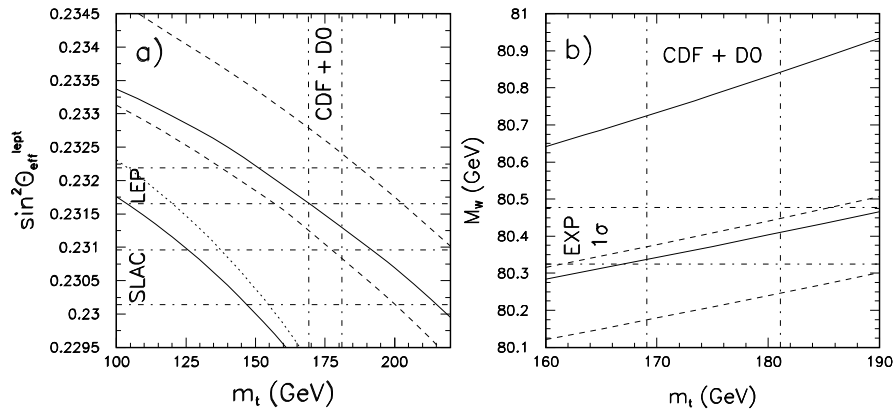,height=5.9cm}
\caption{Predictions for $\sin^2\theta^{lept}_{eff}$ ({\bf a}) and $M_W$ 
({\bf b}) in the SM (the band bounded by the dashed lines) and in the MSSM 
(solid lines) as functions of the top-quark mass. The bands are obtained by
scanning over the MSSM parameters ($M_h$ in the SM) respecting all available
experimental limits. The SLC and the (average) 
LEP measurements for $\sin^2\theta^{lept}_{eff}$ and $1 \sigma$ experimental 
range for $M_W$ are marked by horizontal dash-dotted lines. Dotted line in 
({\bf a}) shows the lower limit for $\sin^2\theta^{lept}_{eff}$ in the MSSM 
if all sparticles are heavier than $Z^0$.}
\label{fig:kane2}
\end{figure}

The important r\^ole 
played in the fit by the precise result for $\sin^2\theta_{eff}^{lept}$ is 
illustrated in Fig. \ref{fig:kane2}a. The world average value (used in 
obtaining the bounds shown in  Fig. \ref{fig:kane1}) is obtained in the SM 
model with $m_t=(175\pm6)$ GeV and $M_h\sim (120-150)$ GeV, with little room 
for additional supersymmetric contribution. Hence, the relevant superpartners 
($\tilde t_L$ and $\tilde b_L$) have to be heavy. With lighter superpartners, 
one obtains the band (solid lines) shown in Fig. \ref{fig:kane2}a. We see that 
the SLD result for $\sin^2\theta_{eff}^{lept}$ leaves much more room for light
superpartners. Thus, settling the SLD/LEP dispute is very relevant for new 
physics. Similar dependence for $M_W$ is shown in Fig. \ref{fig:kane2}b. The 
experimental result, $M_W=80.401\pm0.076$ GeV, is the average of $M_W$ 
measured by UA2, LEP and Tevatron \cite{ROSNER}. 

All squarks of the first two generations as well as sleptons can be still 
at their present lower experimental limits, and the success of the SM in the 
description of the precision electroweak data is still maintained. The same 
applies to the gaugino/higgsino sectors, since they do not give any strong 
$SU_V(2)$ breaking 
effects. In conclusion, most of the superpartners decouple from most of the
electroweak observables, even if very light, ${\cal O}(M_Z)$. This high 
degree of screening follows from the basic structure of the model. The 
remarkable exception is the famous $R_b$ \cite{RB}.

\section{$R_b$ in the MSSM}
As already mentioned, although the new ALEPH measurement \cite{ALEPHRB} of 
$R_b$ is in perfect agreement with its value predicted in the SM, the average 
of all measurements still deviates from the SM value by $\sim1.8\sigma$ 
\cite{LEPEWWG97}. In view of this fact and because the $R_b$ value in the MSSM
is sensitive to different set of parameters than the bulk of the other 
electroweak observables, it is interesting to discuss this observable in more
detail.

In the MSSM there are new contributions to the $Z^0\bar bb$ vertex, namely,
Higgs bosons exchange in the loops, neutralino-sbottom and chargino-stop loops
\cite{BUFI}. For low and intermediate values of $\tan\beta$, the first give 
negative contribution to $R_b$ and to minimize this effect the pseudoscalar 
mass $M_A$ has to be sufficiently large, say, $M_A > {\cal O}(300$ GeV) 
whereas the neutralino-sbottom contribution is negligible. The chargino-stop 
loops can be realized in two ways: with stop coupled to $Z^0$ and with 
charginos coupled to $Z^0$. In both cases the lighter the stop and chargino 
the larger is the positive contribution. 

The Dirac charginos are defined as
\begin{eqnarray}
C^-_i = \left(\matrix{\lambda^-_i\cr\overline\lambda^+_i}\right) ~~~~~~~
i=1,2
\end{eqnarray}
where $\lambda^-_i$ ($\lambda^+_i$) are linear combinations of the negatively
(positively) charged $SU(2)$ gauginos and down-(up-)type higgsinos
\footnote{Up- and down- type higgsinos are superpartners of the Higgs
boson doublets giving masses to the up- and down-type quarks, respectively.}
The $b\tilde t_1 C^-$ coupling is enhanced for a right-handed stop (it is 
then proportional to the top quark Yukawa coupling). Then, however, the stop 
coupling to $Z^0$ is suppressed (it is proportional to $g\sin^2\theta_W$).
Therefore, significant contribution can only come from the diagrams in which 
charginos are coupled to $Z^0$. Their actual magnitude depend on the 
interplay 
of the couplings in the $C^-_i\tilde t_1 b$ vertex and the $Z^0C^-_iC^-_j$ 
vertex. The first one is large only for charginos with large up-higgsino 
component, the second - for charginos with large gaugino component in at least
one of its two-component spinors. It has been observed \cite{CHPO,DPROY}
that, the situation in which both couplings are simultaneously large never 
happens for $\mu>0$. Large $R_b$ can then only be achieved at the expense of 
extremly light $C^-_j$ and $\tilde t_1$. In addition, for fixed $m_{C_1}$ and 
$M_{\tilde t_1}$, $R_b$ is larger for $r\equiv M_2/|\mu| >1$ i.e. for 
higgsino-like chargino as the enhancement of the $C^-_1\tilde t_1 b$ coupling 
is more important than of the $Z^0C^-_1C^-_1$ coupling.

\begin{figure}
\psfig{figure=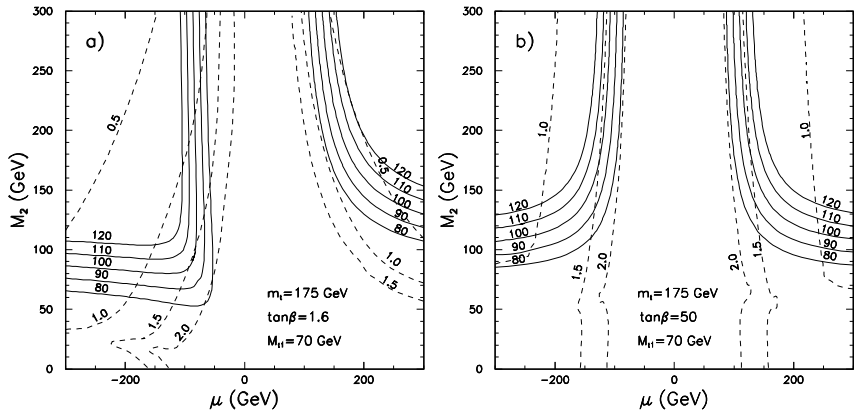,height=6.0cm}
\caption{Contours of constant lighter chargino masses
$m_{C^\pm_1}=$80 $-$ 120 GeV (solid lines) and of 
$\delta R_b\times10^3=$2.0, 1.5, 1.0, 0.5 (dashed lines) in the $(\mu, M_2)$ 
plane for $m_t=175$ GeV, $M_{\tilde t_1} = 70$ GeV. 
{\bf a)}  $\tan\beta=1.6$, {\bf b)}  $\tan\beta=50$ and $M_A=60$ GeV. Chargino 
masses $m_{C^\pm_1}\simlt$83 GeV are already excluded by LEP2.}
\label{fig:kane3}
\end{figure}

For $\mu<0$ the situation is different. In the range $0.5\simlt r\simlt1.5$ 
a light  chargino can be a strongly mixed state with a large up-higgsino 
and gaugino components (the higgsino-gaugino mixing comes from the 
chargino mass matrix). Large couplings in both vertices of the diagram with 
charginos coupled to $Z^0$ give significant increase in $R_b$ even for the 
lighter chargino as heavy as $80-90$ GeV (similar increase in $R_b$ for 
$\mu>0$ requires $m_{C_1}\approx50$ GeV and $M_{\tilde t_1}\approx50$ GeV). 
This is illustrated in Fig. \ref{fig:kane3}a where we show the contours of 
constatnt $\delta R_b$ in the $(M_2,\mu)$ plane for fixed parameters of the
stop sector.

The chargino-stop contributions do not change the value of the left-right 
asymmetry in $b$ quarks ${\cal A}_b\equiv (g^2_L-g^2_R)/(g^2_L+g^2_R)$ 
where $g_L$ ($g_R$) are the effective couplings of left-handed (right-handed) 
$b$ quarks to $Z^0$ (measured at SLD \cite{SLD}) as they mostly modify only 
the left-handed effective coupling $g_L$ \cite{BUFI}. In this case we get 
$\delta {\cal A}_b\approx5.84\times(1-{\cal A}_b^{SM})\times\delta R_b$ i.e.
a very small, positive shift \cite{CLINE}.

An enhancement of $R_b$ is also possible for large $\tan\beta$ values, 
$\tan\beta\approx m_t/m_b$ \cite{BUFI,ROSIEK}. In this case, in addition to 
the stop-chargino contribution (and neutralino-sbottom contribution enhanced 
by large $\tan\beta$) there can be even larger positive contribution from
the $h^0$, $H^0$ and $A^0$ exchanges in the loops, provided those particles 
are sufficiently light (in this range of $\tan\beta$, $M_h\approx M_A$) and 
non-negligible sbottom-neutralino loop contributions. The main difference 
with the low $\tan\beta$ case is the approximate
independence of the results on the sign of 
$\mu$ (which can be traced back to the approximate symmetry of the chargino 
masses and mixings under $\mu\rightarrow -\mu$). Also, the effects are always
maximal for $M_2/|\mu|\gg1$ i.e. for higgsino-like chargino. With present 
experimental constraints ($M_A\simgt60$ GeV \cite{ALEPH}) values of $R_b$ up 
to $\sim0.2178$ 
can still be obtained \cite{CHPO,SOLALTB}. This is illustrated in Fig.
\ref{fig:kane3}b. Significant enhancement of $R_b$ in the large $\tan\beta$ 
regime is, however, rather unlikely as it requires very precise cancellation 
of the SUSY contributions to obtain acceptable rate for 
$b\rightarrow s\gamma$ \cite{CHPO} (see later).

We conclude that additional supersymmetric contributions to the $Z^0\bar{b}b$ 
vertex, from the chargino-right-handed stop loop (and from a light $CP$-odd
Higgs boson for large $\tan\beta$), can be non-negligible when 
both are light, $\sim {\cal O}(M_Z)$. However, even with the chargino and stop
at their present experimental mass limit, the prediction for $R_b$ in the MSSM
depends strongly on the chargino composition (see Fig. \ref{fig:kane3})
and on the stop mixing angle. The values ranging from 
0.2158 (the SM prediction) up to $\sim0.2178$ for both small and large 
$\tan\beta$ values can be obtained (given all the experimental 
constraints) \cite{CHPO,DPROY}. (Only marginal modification of the SM result 
for $R_c$ is possible, though \cite{SOLARC}.) Those predictions hold with or 
without $R$-parity conservation \cite{MYDEB}
and with or without the GUT relation for the gaugino masses. The 
upper bound is reachable for chargino masses up to $\sim90$ GeV provided they 
are mixed gaugino-higgsino states $(M_2/|\mu|\sim1)$ with $\mu<0$ for low 
$\tan\beta$ and higgsino-like for large $\tan\beta$. In the same chargino 
mass range $\delta R_b\approx0$ in the deep higgsino and gaugino regions
for low $\tan\beta$ and gaugino region for large $\tan\beta$. 

In conclusion, the new values of $R_b$ and $R_c$ are good news for 
supersymmetry. At the same time, one should face the fact that, 
unfortunately, in the MSSM
\begin{eqnarray}
\delta R_b^{max}\sim {\cal O}(1 ~\sigma^{exp})\nonumber
\end{eqnarray}
so much better experimental precision is needed for a meaningful discussion. 

\section{$g-2$ and supersymmetry}

One of the best measured electroweak observables is the muon anomalous 
magnetic moment
\begin{eqnarray}
a_\mu =(g_\mu -2)/2=(116 592 300 \pm 840)\times10^{-11}\nonumber
\end{eqnarray}

The theoretical value for $a_\mu$, $a_\mu =(116 591 830\pm150)\times10^{-11}$
is dominated by the ordinary QED contribution (known up to 
${\cal O}(\alpha^5)$) $a^{QED}_\mu =(116 584 706\pm2)\times10^{-11}$
and the hadronic contribution to vacuum polarization 
$a^{had}_\mu =(7020\pm150)\times10^{-11}$ \cite{EID}
\footnote{Recently the 
hadronic photon vacuum polarization contribution has been estimated using the 
ALEPH data for hadronic $\tau$ decays to be $(6950\pm150)\times10^{-11}$
\cite{ALDAHO}.}.
Standard electroweak contribution to $a_\mu$ gives $a^{EW}_\mu =(152\pm
3)\times 10^{-11}$ for the combined one- and two-loop corrections
(the weak 2-loop terms calculated recently \cite{CZAR} are small).
Thus, the present experimental accuracy, $\sim 10^{-3}\%$, 
is sufficient to test only the QED sector of the SM \cite{SEE}. 

A renewed interest in the muon anomalous magnetic moment is due to the ongoing
Brookhaven National Laboratory experiment, with the anticipated accuracy 
$\delta a^{exp}_\mu\approx\pm40\times 10^{-11}$. Even with this measurement 
done with the foreseen accuracy, the one-loop weak corrections can be tested 
only after a substantial reduction of the hadronic vacuum polarization 
uncertainty. This can only be achieved by new measurements of the cross
section for $e^+e^-\rightarrow hadrons$ in the low energy range.
Under the same condition, the precise measurement of $a_\mu$ will be
a very important test of new physics, sensitive to mass scales
beyond the reach of the present accelerators \cite{NATH,ESMA}. 

At present, the requirement that the supersymmetric contribution 
$\delta^{new} a_\mu$ lies within the difference between experimental and 
theoretical results
\begin{eqnarray}
-900\times10^{-11}<\delta^{new} a_\mu < 1900\times10^{-11}\nonumber
\end{eqnarray}
puts already some constraints, though mariginal, on the MSSM parameter space.
In particular, for large $\tan\beta$ the dominant 
supersymmetric contribution due to neutralino-smuon and chargino-sneutrino 
loops gives approximately
\cite{MO,CAGIWA}
\begin{eqnarray}
\delta^{susy} a_\mu\approx\pm{\alpha\over8\pi\sin^2\theta_W}{m^2_\mu\over 
M^2_{SUSY}}\tan\beta\approx\pm150\times10^{-11}
\left({100 {\rm GeV}\over M_{SUSY}}\right)^2\tan\beta
\end{eqnarray}
($M_{SUSY}$ is the average supersymmetric mass and  the sign is correlated 
with the sign of the $\mu$ parameter) and for $\tan\beta\simgt10$ eliminates
some portion of the chargino-sneutrino mass plane \cite{MO,CAGIWA}. It is 
clear that the new Brookhaven experiment if supplemented with the reduction of
the hadronic vacuum polarization uncertainty will enhance the sensitivity
to the chargino-smuon sector of the MSSM.
 
\section{FCNC with light superpartners}

Another important class of processes, where light superpartners could 
manifest themselves through virtual corrections, are the Flavour Changing 
Neutral Current (FCNC) transitions. Gauge invariance, renormalizability and 
particle content of the SM imply the absence (in the lepton sector) or strong 
suppression (in the quark sector) of such processes.  This prediction of 
the SM is in beautiful agreement with the presently available experimental
data. In the MSSM there are two kinds of new contributions 
to the FCNC transitions. First of all, they may originate from flavour mixing 
in the sfermion mass matrices \cite{D83}. The strong suppression of the FCNC 
transisions observed in Nature puts severe constraints on flavour changing 
elements in the sfermion mass matrices \cite{MAGASI}. Even in the absence of
such effects the other kind of new contributions to FCNC processes arise 
through the ordinary $K$-$M$ mixings due to additional exchanges of (light) 
supersymmetric particles in loops.

The present section is devoted to discussing such a scenario.
The only extra MSSM contributions to the FCNC processes we consider are the
charged Higgs boson-top and chargino-stop loops.
The third generation sfermions and chargino(s) are indeed expected to be among 
the lightest superpartners. It is reasonable to assume (as follows from the 
analysis of the renormalization group equations for soft supersymmetry 
breaking parameters) that the first two generations of sfermions are heavier 
and  degenerate in mass. The relevant parameter space is then 
identical to the one tested in corrections to $R_b$ discussed earlier.
Following the results on the precision tests we will assume that the heavier 
stop is heavy enough to decouple and that the lighter one is dominantly 
right-handed, i.e. that stop left-right mixing angle $\theta_{\tilde t}$ 
is relatively small (of order $10^o$). 
	
Within this scenario, sizeable effects can still occur in the neutral meson 
mixing ($K^0$-$\bar K^0$ and $B^0$-$\bar B^0$). Supersymmetric contributions 
to other FCNC processes are either small or screened by long-distance QCD 
effects. A remarkable exception is the inclusive 
weak radiative B meson decay, $B\rightarrow X_s\gamma$, to which light 
superpartners can contribute significantly, and where strong interaction 
effects are under control. Therefore, in the following, we shall focus on 
neutral meson mixing and the $B\rightarrow X_s\gamma$ decay only and summarize
the main points. More extensive discussion is given in ref. \cite{MIPORO}.

With the assumption that the effects of the off diagonal entries 
of the sfermion mass matrices can be neglected, the predictions for 
$\Delta m_{B_d}$ and $\epsilon_K$ can be written as
\begin{eqnarray}
\Delta m_{B_d} &=& \eta_{QCD}  {\alpha_{em}^2 m_t^2 \over 12 
\sin^4\theta_W M_W^4} f_{B_d}^2 B_{B_d} m_{B_d}
\mid V_{tb}V_{td}^\star\mid^2\mid\Delta\mid,
\label{eq:dmbd}\\
\mid\epsilon_K\mid &=&{\sqrt{2}\alpha_{em}^2 m_c^2\over 48 
\sin^4\theta_W M_W^4} f_K^2 B_K 
{m_K\over \Delta m_K}\mid{\cal I}m\Omega\mid,
\label{eq:epsk}
\end{eqnarray}
where $V_{ij}$ are the elements of the $K$-$M$ matrix,
\begin{eqnarray}
\Omega = \eta_{cc}(V_{cs}V_{cd}^\star)^2
+ 2\eta_{ct}(V_{cs}V_{cd}^\star V_{ts}V_{td}^\star)
f\left(\frac{m_c^2}{M_W^2},\frac{m_t^2}{M_W^2}\right)
+\eta_{tt}(V_{ts}V_{td}^\star)^2\frac{m_t^2}{m_c^2}\Delta, 
\label{eq:epskom}
\end{eqnarray}
and
\begin{eqnarray}
f(x,y)=\log{y\over x}+{3y\over4(y-1)}\left(1-{y\over y-1}\log y\right).
\nonumber
\end{eqnarray}

The charged Higgs boson and chargino boxes contribute, together with the SM 
terms, only to the quantity $\Delta$. The QCD correction factors $\eta_{cc}$, 
$\eta_{ct}$, $\eta_{tt}$ and $\eta_{QCD}$ are known up to the next-to-leading 
order \cite{etaNLO}.

The theoretical predictions for $\epsilon_K$ and $\Delta m_{B_d}$ have a well
known uncertainty due to non-perturbative parameters $B_K$, 
$f_{B_d}^2 B_{B_d}$
estimated from lattice calculations. Moreover, the value of the element 
$V_{td}=A\lambda^3(1-\rho-i\eta)$ of the $K$-$M$ matrix (we use the
Wolfenstein parametrization \cite{WOLFENSTEIN}), which is not measured 
directly, is extracted from the fit to the observables
(\ref{eq:dmbd}-\ref{eq:epskom}) and obviously changes after inclusion of 
new, nonstandard, contributions to $\Delta$. Thus, the correct approach is
to fit the parameters $A$, $\rho$, $\eta$ and $\Delta$ in a model independent 
way to the experimental values of $\epsilon_K$ and $\Delta m_{B_d}$ 
\cite{BFZ96,MIPORO} keeping the values of $|V_{cb}|$ and $|V_{ub}/V_{cb}|$ 
within their experimentally allowed ranges
\footnote{The values of $|V_{cb}|$ 
and $|V_{ub}/V_{cb}|$ are known from tree level processes and are practically 
unaffected by new physics which contributes only at one and more loops.}.

\begin{figure}
\psfig{figure=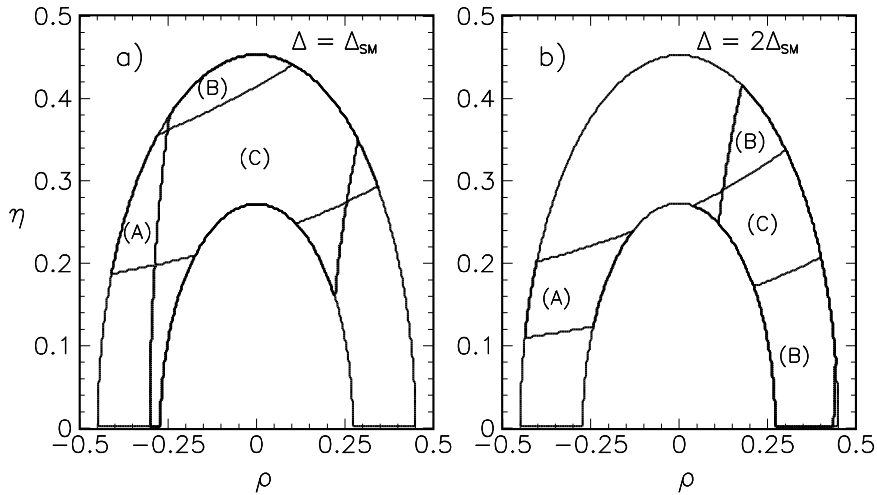,height=6.8cm}
\caption{Allowed regions in the $(\rho, \eta)$ plane for two different
values of $\Delta$. Regions (A) are allowed by $\epsilon_K$, regions (B) - by 
$\Delta m_{B_d}$. Regions (C) are allowed by both measurements 
simultaneously.}
\label{fig:kane4}
\end{figure}

Such a fit with $B_K$ and $f_{B_d}^2 B_{B_d}$ varied in the 
ranges \cite{B96}: $0.6<B_K<0.9$, 
$0.160~{\rm GeV}<\sqrt{f_{B_d}^2 B_{B_d}}<0.240~{\rm GeV}$ gives \cite{MIPORO}
rather liberal ``absolute'' bounds on $\Delta$: $0.2\simlt\Delta\simlt2.0$.
This should be compared with the theoretical prediction for the parameter 
$\Delta$ in the SM: $\Delta=0.53$.  Larger values of $\Delta>\Delta_{SM}$ 
(interesting in the MSSM, as discussed later) prefer $\rho>0$, small values 
of $f_{B_d} (B_{B_d})^{1/2}$ and, to a lesser extent, large $B_K$. For 
instance, $\Delta>1$ requires $\rho>0$ and $f_{B_d} (B_{B_d})^{1/2}<0.19$ GeV.
This is illustrated in  Fig. \ref{fig:kane4}. Values of $\Delta$ smaller than 
$\Delta_{SM}$ prefer negative $\rho$ and large $\eta$. $\Delta\sim\Delta_{SM}$ 
gives the biggest allowed range for $\rho$ and $\eta$ with both $\rho<0$ and 
$\rho>0$ possible.

\begin{figure}
\psfig{figure=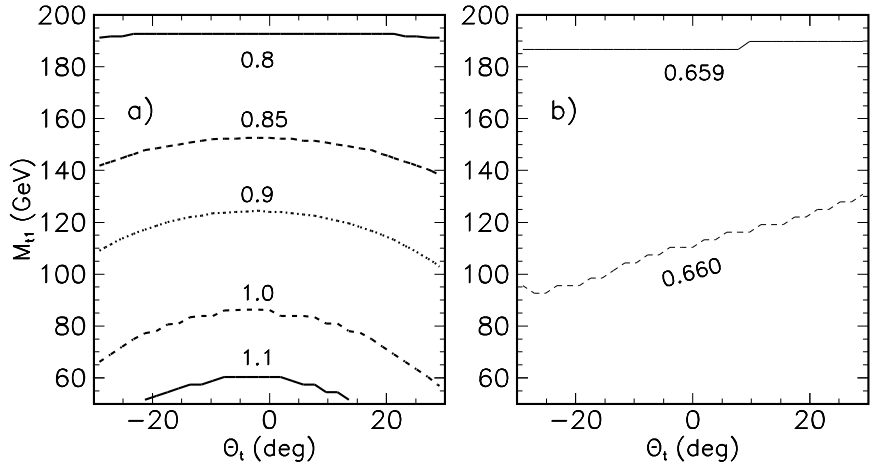,height=6.5cm}
\caption{Contour lines of $\Delta$ as a function of right stop mass and stop 
mixing angle for $\tan\beta=1.8$, $M_{H^+}= 100$ GeV, $M_{\tilde t_2}=250$ 
GeV, $m_{C^\pm}=90$ GeV and $m_t=175$ GeV. {\bf a)} $M_2/\mu=-1$,
{\bf b)} $M_2/\mu=0.1$.}
\label{fig:kane5}
\end{figure}

The model independent bounds for $\Delta$ can be compared with the predictions
for this quantity in the MSSM.
In Fig. \ref{fig:kane5}, we plot contour lines of constant $\Delta$ computed 
in the MSSM with light spectrum, for which SUSY effects are most visible. 
As seen from Fig. \ref{fig:kane5},
the values of $\Delta$ in the MSSM are always bigger than in the SM, i.e. the 
new contributions to $\Delta$ from the Higgs and chargino sectors have the 
same sign as $\Delta_{SM}\approx0.53$ (for $m_t=175$ GeV). This is a general 
conclusion, always 
true for the Higgs contribution and valid also for the chargino-stop 
contribution when SUSY parameters are chosen as specified at the beggining of
this section. The charged Higgs boson contribution increases $\Delta$ by about
0.12 for $M_{H^\pm}=100$ GeV and $\tan\beta=1.8$ as used in Fig. 
\ref{fig:kane5}. The value of the genuine supersymmetric contribution to 
$\Delta$ depends strongly on the ratio $r\equiv M_2/|\mu |$. For small 
values of $r$, when the lighter chargino is predominantly gaugino-like, the 
chargino-stop contribution to $\Delta$ is very small (of order 
$10^{-2}$) and weakly dependent on the lighter stop mass. This can be easily 
understood: In this case, the lighter stop is coupled to the lighter chargino 
mostly through the left-right mixing in the stop sector, and the appropriate 
contribution is suppressed by $\sin^4\theta_{\tilde t}$. For larger values of 
$r$, $r\sim 1$, this contribution is bigger and, due to the interference 
between the diagrams with and without the left-right  mixing, may 
reach its maximal value for $\theta_{\tilde t}\neq 0$, depending on the sign 
of $\mu$. Chargino-lighter stop contribution increases further with 
$M_2/|\mu|$, when lighter chargino is predominantly up-type Higgsino, 
and become again independent on the sign of $\mu$.

Increasing the charged Higgs mass to $M_{H^{\pm}}\approx 500$ GeV and chargino
mass to $m_{C_1^{\pm}}=300$ GeV suppresses the magnitude of each contribution 
by a factor of 3 approximately, but does not change the character of its 
dependence on $\theta_{\tilde t}$. The results illustrated in Fig. 
\ref{fig:kane5} are also weakly dependent on the mass of the left stop: 
Increasing the heavier stop mass, $M_{\tilde t_2}$, from 250 to 500 GeV 
modifies $\Delta$ only marginally.

We conclude that in the $K^0$-$\bar K^0$ and $B^0$-$\bar B^0$ mixing there is 
a room for important supersymmetric contributions. In the presence of such
contributions the values of the $CP$ violation parameters $\rho$ and $\eta$
are different from their SM values and can be tested in the study of the
$CP$ violation in $B$-factories.

We now turn to the discussion of $B\rightarrow X_s\gamma$ decay rate which in
the first approximation is given by simple one-loop graphs. However, the 
strong interaction corrections to these one-loop diagrams are enhanced by the 
large logarithms $\ln(M_W^2/m_b^2)$ and, in the SM, they increase the decay 
rate by a factor of order 2. Thus, resumming these large QCD logarithms up to 
(at least) next-to-leading order (NLO) is necessary to acquire sufficient 
accuracy \cite{BMMP94}. This is 
conveniently done in three steps of which only the first one depends on the 
presence of ``new physics'' (supersymmetry) close to the electroweak scale. 

In the first step one integrates out at the scale 
$Q=M_W$ all heavy fields and introduces the effective Hamiltonian 
\begin{eqnarray}
H_{eff}=-{4G_F\over\sqrt2} V^*_{ts}V_{tb}\sum_{i=1}^8 C_i(Q){\cal O}_i(Q) 
\label{heff}
\end{eqnarray}
where ${\cal O}_i$ are the operators and $C_i(Q)$ are their Wilson 
coefficients. The relevant for $B\rightarrow X_s\gamma$  operators are
\begin{eqnarray}
{\cal O}_7 &=&{e\over16\pi^2} m_b (\bar{s}_L\sigma^{\mu\nu} b_R) F_{\mu\nu} 
\label{P7} \\  
{\cal O}_8 &=&{g_s\over16\pi^2} m_b (\bar{s}_L\sigma^{\mu\nu} T^a b_R) 
G_{\mu\nu}^a \label{P8}
\end{eqnarray}
where $F_{\mu\nu}$ and $G_{\mu\nu}^a$ are the photonic and gluonic field 
strength tensors, respectively. The leading-order SM \cite{INLI} and MSSM 
\cite{BEBOMARI} contributions to the coefficients $C_7(M_W)$ and 
$C_8(M_W)$ are well known. The next-to-leading corrections to $C_7(M_W)$ have 
been computed fully only in the case case of the SM \cite{AY94}. In the 
supersymmetric case, only contributions proportional to logarithms of 
superpartner masses are known \cite{A94}. 

In the next step, resummation of large logarithms $\ln(M_W^2/m_b^2)$ is 
achieved by evolving the coefficients $C_i(Q)$ from $Q\sim M_W$ to $Q\sim m_b$
according to the renormalization group equations. The necessary for the 
complete NLO evolution coefficients of the RGE have been computed only 
recently \cite{CMM96}.

Finally, the Feynman rules derived from the effective Hamiltonian at the scale
$Q\sim m_b$ are used to calculate the $b$-quark decay rate 
$\Gamma(b\rightarrow X_s\gamma)$ which is a good approximation to the 
corresponding $B$-meson decay rate \cite{FLS94}. Radiative corrections to 
this computation, necessary to achieve NLO precision of the whole procedure, 
have also been computed recently \cite{GHW96}.

\begin{figure}[htbp]
\begin{center}
\begin{tabular}{p{0.48\linewidth}p{0.48\linewidth}}
\mbox{\epsfig{file=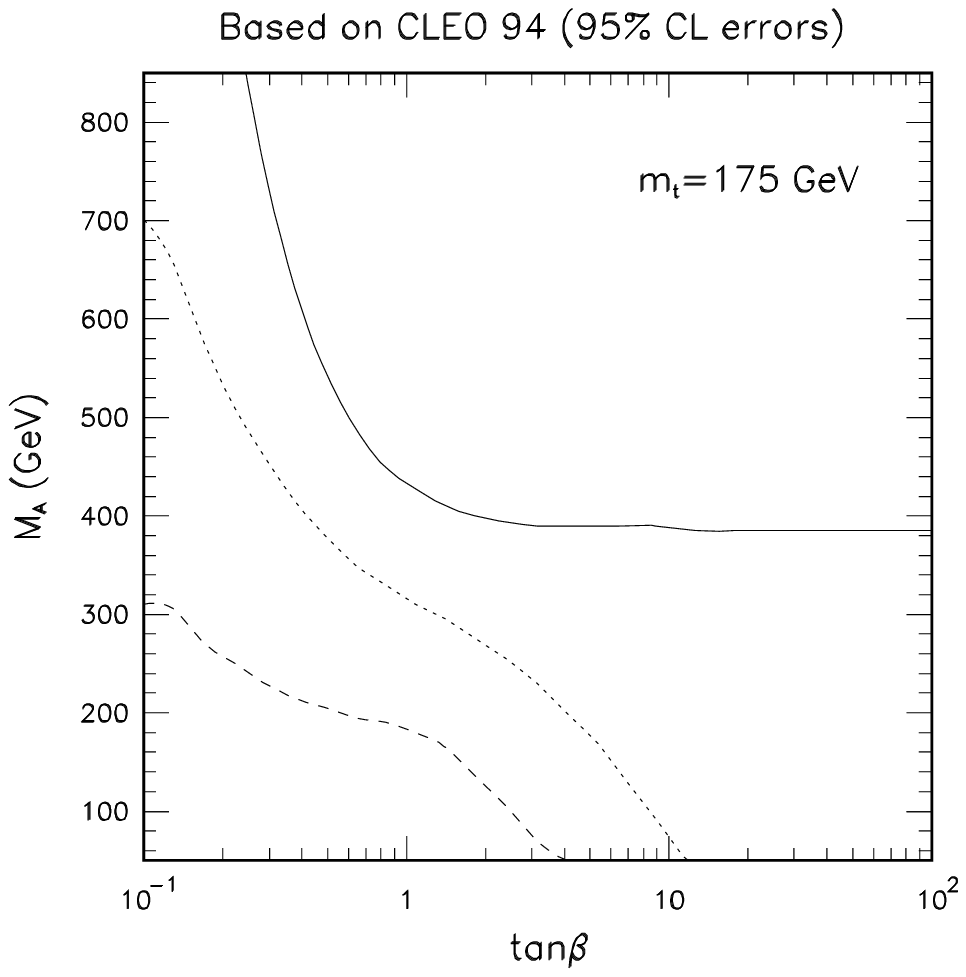,width=\linewidth}}
&
\mbox{\epsfig{file=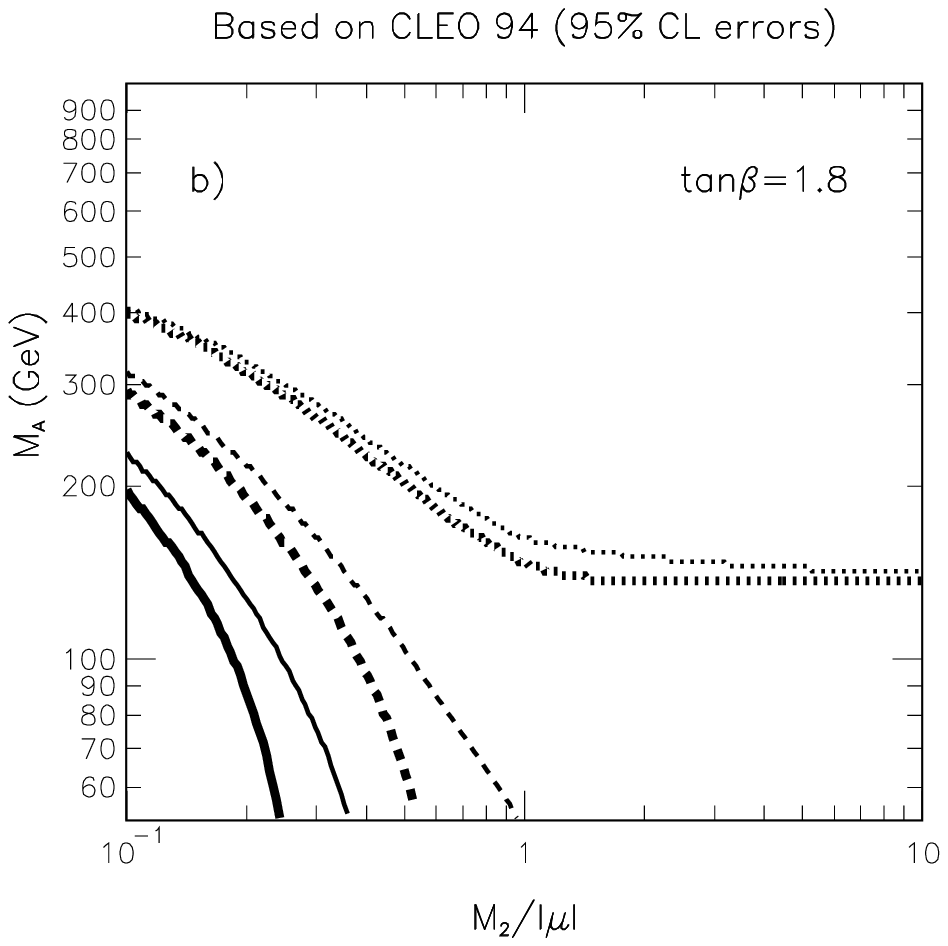,width=\linewidth}}
\\
\end{tabular}
\caption{{\bf a)} Lower limits on $M_A$ from $b\rightarrow s\gamma$ 
as a function of $\tan\beta$.
Solid line correspond to very heavy, $>{\cal O}(1$ TeV) sparticles.
Dashed (dotted) line show the limit for $m_{C_1}=M_{\tilde t_1}=$250 (500) 
GeV. {\bf b)} Lower limits on $M_A$ as a function of $M_2/|\mu|$,
based on CLEO $BR(B\rightarrow X_s\gamma)$ measurement. Thick lines
show limits for $\mu>0$, thin lines for $\mu<0$. Solid, dashed and
dotted lines show limits for lighter stop and chargino masses
$M_{\tilde t_1}=m_{C_1^\pm}=90$, 150 and 300 GeV,
respectively.\label{fig:kane67}}
\end{center}
\end{figure}

We stress again that all these calculations are identical in the SM and MSSM 
except for the initial numerical values of the Wilson coefficients $C_7$ and 
$C_8$ at $Q\sim M_W$ which contain the information about ``new physics''. 
Another important remark is that even in the NLO computation, the theoretical
prediction for $\Gamma(b\rightarrow X_s\gamma)$ still has an uncertainy 
(shown in Fig \ref{fig:kane67}a) of order 15\% \cite{CMM96} which is always 
taken into account in the bounds on sparticle masses presented below.

Fig. \ref{fig:kane67}a shows the lower limits on the mass of the 
$CP$-odd MSSM Higgs boson mass \footnote{The charged Higgs
boson mass is in one-to-one correspondence with $M_A$: 
$M_{H^\pm}^2=M_A^2+M_W^2$ (up to small radiative corrections \cite{MYMH}).},
$M_A$, as a function of $\tan\beta$. Solid lines correspond to the 
case when all the superpartner masses are very large (above 1 TeV). In this 
case, the MSSM results are the same as in the Two Higgs Doublet Model (2HDM). 
Dashed (dotted) lines in Fig. \ref{fig:kane67}a show the same limits 
in the presence of chargino and stop with masses $m_{C_1}=M_{\tilde t_1}=500$
(250) GeV (with all other sparticles heavy) obtained by scanning over the 
values of $r=M_2/\mu$ and $\theta_{\tilde t}$. In the presence of light stop 
and chargino limits on $M_A$ are significantly weaker and totally disappear 
for large values of $\tan\beta$ for which the chargino-stop contribution can 
be very large. The price however is a high degree of fine tuning
(see the next Section).

The existing measurement of $BR(b\rightarrow s\gamma)$ imposes already 
significant constraints on the MSSM parameter space. To understand them, 
it is important to remember that the charged Higgs contribution to the 
$b\rightarrow s\gamma$ amplitude has always the same sign as the SM one 
whereas the chargino-stop contribution to this amplitude may have opposite 
sign. Since the actually measured value of $BR(b\rightarrow s\gamma)$ is 
close to the SM prediction, SUSY and charged Higgs contributions must either 
be small by themselves or cancel each other to a large extent. There exists, 
however, a third possibility where negative chargino-stop contribution 
overcomes the SM and charged Higgs ones yielding the correct absolute magnitude
of the total amplitude but with the opposite sign compared to the SM case.
In paticular, it is worth stressing that the supersymmetric contribution, 
coming from a light chargino and stop, can provide a natural mechanism for 
lowering the $b\rightarrow s\gamma$ rate compared to the SM value, in 
agreement with the trend seen in the present data \cite{BEBOMARI,CHPO}.

Another important observation is that, large chargino-stop contribution to 
$b\rightarrow s\gamma$ amplitude arise when the chargino is 
higgsino-like rather then gaugino-like i.e. when $M_2/|\mu | >1$. 
In addition, the size of the chargino-stop contribution can be modified by 
changing the stop mixing angle $\theta_{\tilde t}$.  

\begin{figure}[htbp]
\begin{center}
\begin{tabular}{p{0.48\linewidth}p{0.48\linewidth}}
\mbox{\epsfig{file=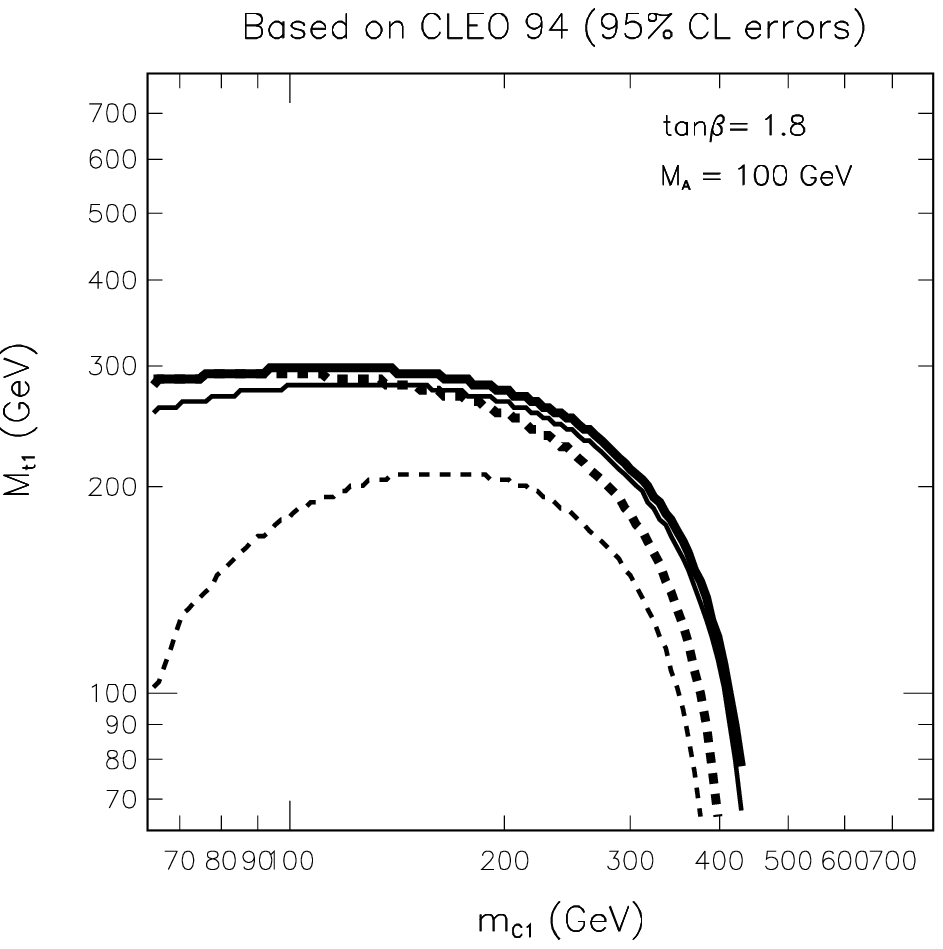,width=\linewidth}}
&
\mbox{\epsfig{file=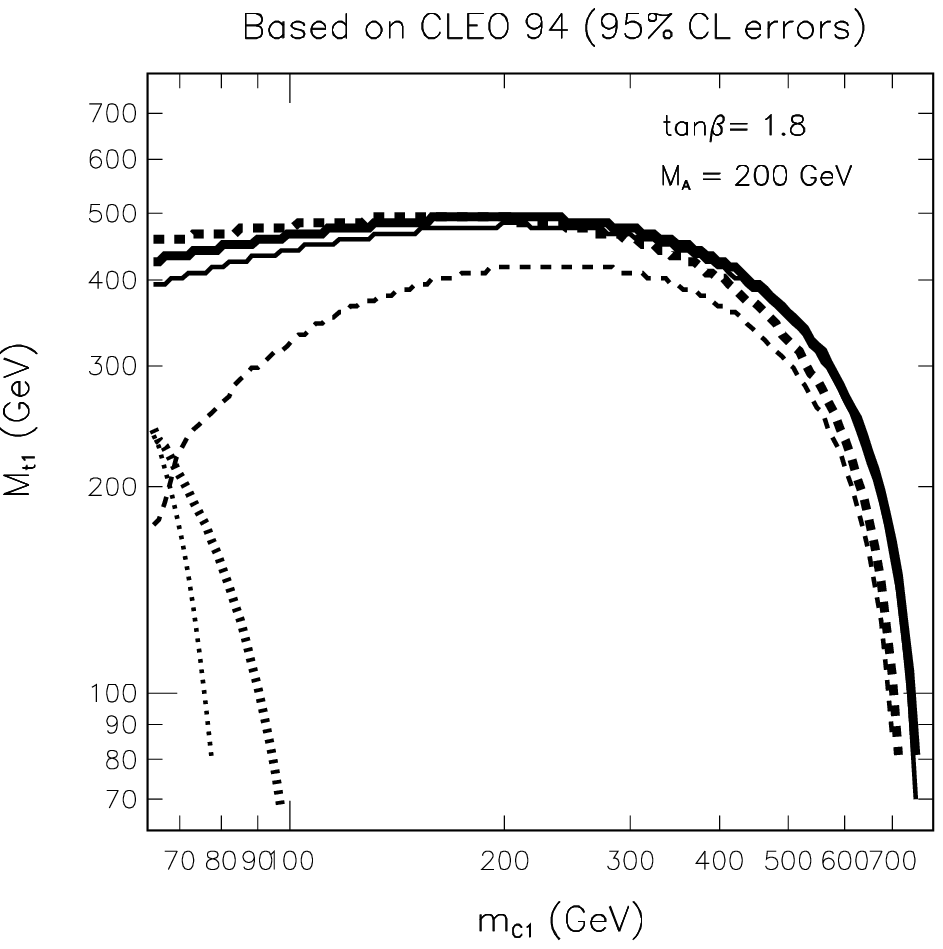,width=\linewidth}}
\\
\end{tabular}
\caption{Bounds on $(M_{\tilde{t_1}}, m_{C_1^\pm})$ plane
for $\tan\beta=1.8$ and $M_A=100$ and 200 GeV. Thick lines show limits
for $\mu>0$, thin lines for $\mu<0$. Dotted, dashed and solid lines
show limits for $M_2/|\mu|=0.1$, 1 and 10, respectively.
\label{fig:kane9}}
\end{center}
\end{figure}

Fig. \ref{fig:kane67}b (taken from ref. \cite{MIPORO}) shows the lower limit 
on the allowed pseudoscalar Higgs boson mass $M_A$ as a function of 
$r=M_2/|\mu|$ for three different values of the lighter chargino and lighter 
stop masses. In Fig. \ref{fig:kane9}, the limits on lighter chargino and 
lighter stop mass for chosen $M_A$ and $M_2/\mu$ values is plotted. In both 
plots a scan over $\theta_{\tilde t}$ in the range 
$-60^{\circ}<\theta_{\tilde t}<60^{\circ}$ has been performed.

Fig. \ref{fig:kane67}b shows that for small $M_2/|\mu|$, i.e. for gaugino-like
lighter chargino (when the chargino-stop contribution to 
$BR(b\rightarrow s\gamma)$ is suppressed) the resulting limits on $M_A$ are 
quite strong even for very light chargino and stop e.g.
$M_A\geq{\cal O}(200$ GeV) for $M_{\tilde t_1}=m_{C_1^\pm}=90$ GeV 
(we take 95\% errors of CLEO measurement). The limits decrease when 
$M_2/|\mu|$ increases and approximately saturate for $M_2/|\mu|\geq 1$. 
Similar effects are visible in Fig. \ref{fig:kane9} where {\em upper} bounds 
on $M_{\tilde t_1}$ are shown as a function of the lighter chargino mass
$m_{C_1^\pm})$. For small $M_2/|\mu|$, very light stop and chargino are 
necessary to cancel the charged Higgs contribution. Thus, the corresponding 
upper limits on their masses are very strong. For large $M_2/|\mu|$, chargino 
and stop even 2-3 times heavier than the charged Higgs are allowed.

\section{Large effects for $\tan\beta\sim m_t/m_b$}

Large values of $\tan\beta$, $\tan\beta\sim m_t/m_b$, have been frequently 
advocated in the literature as a possible dynamical explanation of the
large  top to bottom quark  mass ratio \cite{BA}. In this 
scenario the Yukawa couplings of the down-type quarks and leptons 
are enhanced leading in some cases to large loop effects. 
One example of this kind is the already discussed large contribution of the
chargino-sneutrino loop to $g-2$ of the muon.

\begin{figure}
\center
\psfig{figure=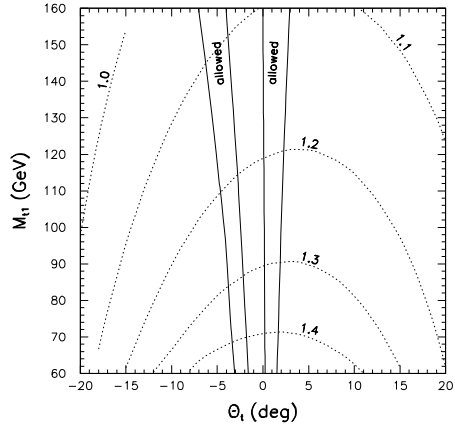,height=6.0cm}
\caption{Allowed by the CLEO result regions in the plane 
$(\theta_{\tilde t}, M_{\tilde t_1})$ for $\tan\beta=50$, $M_A=60$ GeV 
($M_{H^\pm}\approx100$ GeV), $m_{C_1}=90$ GeV, and $M_2/\mu=1.5$. Contours of 
constant $\delta R_b\times10^3$ are also shown.}
\label{fig:kane10}
\end{figure}

Particularily large in this regime are the chargino-stop corrections
to the $b\rightarrow s\gamma$ amplitude. Indeed, in the limit of higgsino-like
lighter chargino we get 
\begin{eqnarray}
C_7^{C_1\tilde t_1}(M_W)\approx
-{m^2_t\over2m^2_{C_1}}\cos^2\theta_{\tilde t_1}f^{(1)}_\gamma(x)
\pm\tan\beta{m_t\over2m_{C_1}}\sin\theta_{\tilde t}\cos\theta_{\tilde t}
f^{(3)}_\gamma(x)
\label{eqn:bsgcst}
\end{eqnarray}
where $x\equiv(M_{\tilde t_1}/m_{C_1})^2$, functions $f^{(i)}_\gamma(x)$ are
defined in the second paper of ref. \cite{BEBOMARI} and the sign in the second
term is the same as the sign of the $\mu$ parameter. It is clear that for 
large
$\tan\beta$, $m_{C_1}\sim M_W$ and the stop mixing angle not too small the 
second term dominates and is much larger than the SM $W^\pm$-$t$ contribution 
\begin{eqnarray}
C_7^{tW^\pm}= {3\over2}{m^2_t\over M_W^2}
f^{(1)}_\gamma\left({m^2_t\over M_W^2}\right)
\end{eqnarray}
Moreover, the higgsino-like chargino-stop contribution vanishes only as 
$1/m_{C_1}$ and remains nonegligible up to relatively large $m_{C_1}$. For 
the gaugino-like higgsino, instead, we get:
\begin{eqnarray}
C_7^{C_1\tilde t_1}(M_W)\approx 
-{M^2_W\over m^2_{C_1}}\cos^2\theta_{\tilde t}f^{(1)}_\gamma(x)
\end{eqnarray}
which is much smaller and vanishes as $1/m^2_{C_1}$. Large contribution from 
light higgsino-like chargino and stop can be cancelled by the charged Higgs 
boson loop. However, this requires a high degree of fine-tuning. 
This is illustrated in Fig. \ref{fig:kane10} where for fixed chargino
parameters and charged Higgs boson mass the allowed region in the plane 
$(\theta_{\tilde t}, M_{\tilde t_1})$ consists of two very narrow bands 
(corresponding to two different signs of the total amplitude). Outside these
bands the generic prediction for the $BR(b\rightarrow s\gamma)$ is one-two 
orders of magnitude larger than the value measured by CLEO.

For similar reasons, the existence of very light, ${\cal O}(M_Z)$, 
pseudoscalar and charged Higgs bosons
(and consequently significant enhancement of $R_b$) in the large $\tan\beta$
regime is rather unlikely. Indeed, charged Higgs contribution by itself would
then give too high a rate for $b\rightarrow s\gamma$. It can be compensated
by the chargino-stop loop but, again, at the expense of strong fine-tuning
(due to the $\tan\beta$ enhancement factor present in eq. (\ref{eqn:bsgcst}))
Nothing, of course, prevents the existence of light gaugino-like chargino and 
stop in the large $\tan\beta$ regime. For heavy enough pseudoscalar $A$ and 
chargino or stop (or both) the contribution of eq. (\ref{eqn:bsgcst}) may have,
however, interesting consequences. Indeed, for negative values of 
$A_t\times\mu$ the rate can be easily smaller than the one of the SM 
\cite{CAOLPOWA} in agreement with the trend of the data. (This scenario can 
be realized in supergravity models with non-universal soft terms \cite{BOOLPO}
and in the gauge mediated models with $B=0$ \cite{RASA}).

Another class of interesting effects in the large $\tan\beta$ regime 
originate from finite corrections to the down-type quark and lepton Yukawa 
couplings \cite{HARASA}. For sparticle masses $\simgt M_Z$ these effects can 
be concisely described by the effective lagrangian \cite{HEMPF_PRIV}
\begin{eqnarray}
{\cal L}_{Yuk} &=&-Y_{1ab}^dH_1\bar d_{Ra}q_{Lb} 
               - Y_{2ab}^dH_2^\dagger\bar d_{Ra}q_{Lb}\nonumber\\
               &-& Y_{2ab}^uH_2\bar u_{Ra}q_{Lb} 
               - Y_{1ab}^uH_1^\dagger\bar u_{Ra}q_{Lb} + h.c.
\label{eqn:efflagr}
\end{eqnarray}
(we suppressed the terms with lepton interactions) describing Yukawa 
interactions arising after integrating out (heavy) sparticles \cite{HEMPF}. 
At the tree level, terms with $Y_{2ab}^d$ and $Y_{1ab}^u$ are absent in the
MSSM (and in the SM). They are, however, generated by triangle diagrams (with 
helicity flips on fermion lines) with squarks and either gauginos or higgsinos
circulating in loops \footnote{These diagrams are finite due to the so-called 
non-renormalization theorems which in the case of unbroken supersymmetry 
would also force these corrections to vanish.}. Most interesting effects 
are due to the new term $Y_{2ab}^d$ which reads 
\begin{eqnarray}
Y_{2ab}^d&=&Y_{1ae}^d\left(\delta_{eb}
         {2\over3}{\alpha_s\over\pi}\mu m_{\tilde g}
         I(M^2_{\tilde q_L},M^2_{\tilde d_R},m^2_{\tilde g})\right.\nonumber\\
         &+&\left.{1\over(4\pi)^2}(Y_2^{u\dagger}Y_2^u)_{eb}\mu A_{\tilde u}
         I(M^2_{\tilde q_L},M^2_{\tilde u_R},\mu^2)\right)
         +{\rm smaller ~terms}\label{eqn:pophemp}\\
         &=&Y_{1ae}^d\Delta_{eb}\nonumber
\end{eqnarray}
where $A_{\tilde u}$ are the trilinear soft supersymmetry breaking terms (for 
simplicity we assume that the soft SUSY breaking matrices $M^2_{\tilde q_L}$, 
$M^2_{\tilde d_L}$ and $A_{\tilde u}$ are all proportional to the unit matrix)
and 
\begin{eqnarray}
I(x,y,z)\equiv{xy\log(x/y)+yz\log(y/z)+zx\log(z/x)\over(x-y)(y-z)(z-x)}
\nonumber
\end{eqnarray}

The presence of $Y_{2ab}^d$ modifies the value of $Y_{1ab}^d$ and (neglecting
nondiagonal terms in Yukawa couplings) we get
\begin{eqnarray}
Y_b\equiv Y_{1bb}^d = 
{e\over\sqrt2s_W}{m_b\over M_W}{\sqrt{1+\tan^2\beta}\over
1+\tan\beta\Delta_{bb}}
\end{eqnarray}
It is clear that sizeable effects appear for large values of $\tan\beta$ and 
light sparticles involved. Moreover, these effects vanish only as 
$1/M_{soft}$ and persist therefore even for relatively heavy sparticles. The
corrections affect all processes where the bottom quark Yukawa coupling is 
involved. In particular, they modify the well known limits on the 
$(\tan\beta, M_{H^\pm})$ plane \cite{KRPO} derived from the experimental 
result for $b\rightarrow c\tau\bar\nu_\tau$ \cite{SOLABCTAU}. Qualitatively,
with these corrections included, $Y_b$ becomes larger for $\mu<0$ and 
enhaces the contribution of the $H^+$ Higgs boson to the process 
$b\rightarrow c\tau\bar\nu_\tau$ strenghtening thus the limits on the 
$(\tan\beta, M_{H^\pm})$ plane. For $\mu>0$, instead, the bound is weakened
and can even dissapear. For detailed discussion, 
see the ref. \cite{SOLABCTAU}.
 
The same effective lagrangian (\ref{eqn:efflagr}) describes also the dominant
part of the (large) supersymmetric corrections to the processes
$H^+\rightarrow tb$ \cite{SOLAHTB} and  
$t\rightarrow H^+b$ which in the case of large $\tan\beta$ and sufficiently 
light $H^+$ (e.g. when $R_b$ is enhanced) competes with the standard decay
$t\rightarrow W^+b$ \cite{SOLATHB}.

These corrections, interpreted as supersymmetric threshold corrections, are 
also very important in context of the GUT models and unification of the 
Yukawa couplings.  In particular, they significantly lower the values of  
$\tan\beta$ and $m_t$ predicted from the bottom-tau Yukawa coupling 
unification (for details see ref. \cite{HARASA,CAOLPOWA})
and also, when their full generation dependence is taken into account,
they significantly modify the naive predictions of the GUT models
for CKM mixing angles \cite{BLPORA}.

\section{Summary}

There is an apparent contradiction between the hierarchy problem (which 
suggest
new physics to be close to the electroweak scale) and the striking success of
the Standard Model in describing the electroweak data. The supersymmetric 
extension of the SM offers an interesting solution to this puzzle. The bulk 
of the electroweak data is well screened from supersymmetric loop effects, 
due to the structure of the theory, even with superpartners generically light,
${\cal O}(M_Z)$. The only exception are the left-handed squarks of the third 
generation which have to be $\simgt {\cal O}(300$ GeV) to maintain the success
of the SM. The other superpartners can still be light, at their present 
experimental mass limits, and would manifest themselves through virtual 
corrections to the small number of observables such as $R_b$, 
$b\rightarrow s\gamma$, $K^0$-$\bar K^0$ and  $B^0$-$\bar B^0$ mixing and a 
few more for large $\tan\beta$. Those effects are very interesting but require
still higher experimental precision to be detectable. 

Our goal here was to study unconstrained minimal supersymmetric model, with
arbitrary soft supersymmetry breaking parameters. Under stronger assumption,
e.g. of universal soft terms at the GUT scale \cite{CMSSM,HARASA,CAOLPOWA}, 
one can get from virtual effects stronger
constraints on the superpartner spectrum (for recent studies see e.g.
refs. \cite{WAGN}).

\section*{Acknowledgments}
We  would like to thank  M. Misiak and J. Rosiek for their contribution
to this review.

This work was supported by Polish State Commitee for Scientific Research under
grant 2 P03B 040 12 (for 1997-98). The work of PHCh was also partly supported 
by the U.S.-Polish Maria Sk\l odowska-Curie Joint Fund II (MEN/DOE-96-264).


\begin{thebibliography}{99}

\bibitem{APCA} T. Appelquist, J. Carazzone {\sl Phys. Rev.} {\bf D11} (1976)
               2856.

\bibitem{ELLISY} J. Ellis, G. Fogli, E. Lisi {\sl Phys. Lett.} {\bf 292B}
                 (1992) 427, {\bf 318B} (1993) 375, {\bf 333B} (1994) 118,
                 P.H. Chankowski, S. Pokorski {\sl Phys. Lett.} {\bf 356B}
                 (1995) 307.

\bibitem{ALBA} G. Altarelli, R. Barbieri {\sl Phys. Lett.} {\bf 253B} (1991) 
               161;
               G. Altarelli, R. Barbieri, S. Jadach {\sl Nucl. Phys.} 
               {\bf B369} (1993) 3;
               G. Altarelli, R. Barbieri, F. Caravaglios  {\sl Phys. Lett.} 
               {\bf 349} (1995) 145, {\sl Nucl. Phys.} {\bf B405} (1993) 3.
               
\bibitem{BLON} A. Blondel in {\sl Proceedings of the
                  XXVIII Int. Conf. on High Energy Physics} Warsaw, 1996, Z.
                  Ajduk, A.K. Wr\'oblewski eds., World Scientific Singapore.

\bibitem{LEPEWWG97} The LEP Electroweak Working Group, CERN preprint 
                    LEPEWWG/97-01.

\bibitem{EID} S. Eidelman, F. Jegerlehner, {\sl Z. Phys.} {\bf C67} (1995) 
              585.

\bibitem{ELLIS} J. Ellis, G. Fogli, E. Lisi {\sl Phys. Lett.} {\bf 389B}
                (1996) 321.

\bibitem{ABC} G. Altarelli, R. Barbieri, F. Caravaglios {\sl Phys. Lett.} 
              {\bf 314B} (1993) 357.
                 
\bibitem{ELFOLI} J. Ellis, G. Fogli, E. Lisi {\sl Phys. Lett.} {\bf 324B}
                 (1994) 173.

\bibitem{CP96.PL} P.H. Chankowski, S. Pokorski {\sl Phys. Lett.} {\bf 366B}
                  (1996) 188.

\bibitem{DABFIT} W. de Boer, A. Dabelstein, W.F.L. Hollik, W. M\"osle 
                 preprint IEKP-KA-96-08 (hep-ph/9609209).

\bibitem{STU} M.E. Peskin, T. Takeuchi {\sl Phys. Rev. Lett.} {\bf 65} 
              (1990) 964, {\sl Phys. Rev.} {\bf D46} (1992) 381;

\bibitem{JAPP} T. Inami, C.S. Lim, A. Yamada {\sl Mod. Phys. Lett.} {\bf A7}
              (1992) 2789.

\bibitem{MYDR} D. Garcia, J. Sol\`a {\sl Mod. Phys. Lett.} {\bf A9} 
               (1994) 211;
               P.H. Chankowski et al. {\sl Nucl. Phys.} {\bf B417} (1994) 101.

\bibitem{BACAFR} R. Barbieri, M. Frigeni, F. Caravaglios {\sl Phys. Lett.} 
                 {\bf 279B} (1992) 169.

\bibitem{HABER} R. Barbieri, M. Frigeni, F. Giuliani, H.E. Haber 
                {\sl Nucl. Phys.} {\bf B341} (1990) 309;
                H.E. Haber in {\sl Proc. of the Workshop on the INFN 
                Eloisotron Project} (hep-ph/9305248).

\bibitem{ROSNER} J. L. Rosner preprint EFI-97-18 (hep-ph/9704331).

\bibitem{CLINE} P. Bamert et al. {\sl Phys. Rev.} {\bf D54} (1996) 4275.

\bibitem{ALEPHRB} I. Tomalin (ALEPH Collaboration) in {\sl Proceedings of the
                  XXVIII Int. Conf. on High Energy Physics} Warsaw, 1996, Z.
                  Ajduk, A.K. Wr\'oblewski eds., World Scientific Singapore.

\bibitem{BUFI} M. Boulware, D. Finnell {\sl Phys. Rev.} {\bf D44} (1991) 2054.

\bibitem{ROSIEK} J. Rosiek {\sl Phys. Lett.} {\bf 252B} (1990) 135;
                 A. Denner et al. {\sl Z. Phys.} {\bf C51} (1991) 695.

\bibitem{RB}  G.L. Kane C. Kolda J.D. Wells
               {\sl Phys. Lett.} {\bf338B } (1994) 219;
               P.H. Chankowski, S. Pokorski in Proceedings of the 
               Beyond the Standard Model IV conference, 
               Lake Tahoe C.A., eds. J.F. Gunion, T. Han, J. Ohnemus
               (1994) p. 233;
               D. Garcia, R. Jim\'enez J. Sol\`a {\sl Phys. Lett.} {\bf347} 
               (1995) 309, 321, E {\bf351B} (1995) 602;
               D. Garcia, J. Sol\`a {\sl Phys. Lett.} {\bf354B} (1995) 335,
               {\bf357B} (1995) 349;
               G.L. Kane R.G. Stuart, J.D. Wells
               {\sl Phys. Lett.} {\bf 354B} (1995) 350; 
               P.H. Chankowski, S. Pokorski {\sl Phys. Lett.} {\bf 366B} 
               (1996) 188;
               G.L. Kane, J.D. Wells {\sl Phys. Rev. Lett.} {\bf 76}
               (1996) 869;
               J. Ellis, J.L. Lopez, D.V. Nanopoulos {\sl Phys. Lett.} 
               {\bf 372B} (1996) 95, {\bf 397B} (1997) 88.

\bibitem{CHPO} P.H. Chankowski, S. Pokorski {\sl Nucl. Phys.} {\bf B475}
               (1996) 3.

\bibitem{DPROY} M. Drees et al. {\sl Phys. Rev.} {\bf D54} (1996) 5598.

\bibitem{SOLALTB} J. Sol\`a, {\sl Phys. Lett.} 
                  {\bf 357B} (1996) 349.

\bibitem{SOLARC} D. Garcia, J. Sol\`a, {\sl Phys. Lett.} 
                 {\bf 354B} (1995) 335. 

\bibitem{ALEPH} The ALEPH Collaboration preprint CERN PPE/97-071

\bibitem{MYDEB} P.H. Chankowski, D. Choudhury, S. Pokorski {\sl Phys. Lett.} 
                {\bf 389B} (1996) 677;
                P.H. Chankowski, S. Pokorski {\sl Acta Phys. Pol.} 
                {\bf B27} (1996) 1719.

\bibitem{SLD} P. Rowson (SLD Collaboration) to appear in {\sl Proceedings of 
              the XXXII Rencontres de Moriond, Electroweak Session},
              J. Tr\^an Van Than ed., Edition Frontieres.

\bibitem{SEE} See e.g. {\sl Proceedings of the 10th International Symposium on
              High Energy Spin Physics}, Nagoya, Japan 1992; {\sl Proceedings
              of the International Symposium on the Future of Muon Physics},
              Heidelberg (1991), edited by K. Jungmann,
              V.W. Hughes and D. zu Putliz [{\sl J. Phys.} {\bf C56} (1991)].

\bibitem{CZAR} A. Czarnecki, B. Krause, W. Marciano, 
               {\sl Phys. Rev.} {\bf D52} (1995), 2619;
               S. Peris, M. Perrottet, E. de Rafael, {\sl Phys. Lett.} 
               {\bf B355} (1995) 523.

\bibitem{ALDAHO} R. Alemany, M. Davier, A. H\"ocker, preprint LAL 97-02 
                 (hep-ph/9703220).


\bibitem{NATH} P. Nath in {\sl Proceedings of the
               XXVIII Int. Conf. on High Energy Physics} Warsaw, 1996, Z.
               Ajduk, A.K. Wr\'oblewski eds., World Scientific Singapore;
               M. Krawczyk, J. \.Zochowski {\sl Phys. Rev.} {\bf D55} (1997) 
               6968.

\bibitem{ESMA} R. Escribano, E. Masso preprint UAB-FT-385, 1996  
               (hep-ph/9607218).

\bibitem{MO} J. Lopez, D.V. Nanopoulos, X. Wang {\sl Phys. Rev.} {\bf D49}
             (1991) 366;
             U. Chattopadhyay, P. Nath {\sl Phys. Rev.} {\bf D53} (1996) 1648;
	     T. Moroi {\sl Phys. Rev.} {\bf D53} (1996) 6565

\bibitem{CAGIWA} M. Carena, G.-F. Giudice, C.E.M. Wagner {\sl Phys. Lett.}
               {\bf 390B} (1997) 234. 

\bibitem{D83} M.J. Duncan {\sl Nucl. Phys.} {\bf B221} (1983) 285;
              J.F. Donoghue, H.-P. Nilles, D. Wyler {\sl Phys. Lett.} 
              {\bf 128B} (1983) 55;
              A. Boquet, J. Kaplan, C.A. Savoy {\sl Phys. Lett.} {\bf 148B}
              (1984) 69.

\bibitem{MAGASI} A. Masiero, F. Gabbiani
                 {\sl Nucl. Phys.} {\bf B322} (1989) 235;
                 J.S. Hagelin, S. Kelley, T. Tanaka {\sl Nucl. Phys.} 
                 {\bf B415} (1994) 293;
                 A. Masiero, F. Gabbiani, L. Silvestrini 
                 {\sl Nucl. Phys.} {\bf B477} (1996) 321.

\bibitem{etaNLO} A.J. Buras, M. Jamin, P.H. Weisz {\sl Nucl. Phys.} 
                 {\bf B347} (1990) 491;
                 S. Herrlich, U. Nierste {\sl Nucl. Phys.} {\bf B419}
                 (1994) 292, {\sl Phys. Rev.} {\bf D52} (1995) 6505.

\bibitem{WOLFENSTEIN} L. Wolfenstein {\sl Phys. Rev. Lett.} 
                      {\bf 51} (1983) 1945.

\bibitem{BFZ96} G.C. Branco, G.C. Cho, Y. Kizukuri, N. Oshimo 
	        {\sl Phys. Lett.} {\bf 337B} (1994) 316, 
                {\sl Nucl. Phys.} {\bf B449} (1995) 483;
                G.C. Branco, W. Grimus, L. Lavoura
                {\sl Phys. Lett.} {\bf 380B} (1996) 119;
                A. Brignole, F. Feruglio, F. Zwirner 
        	{\sl Zeit. Phys.} {\bf C71} (1996) 679;

\bibitem{MIPORO} M. Misiak, S. Pokorski, J. Rosiek preprint IFT 3/97
                (hep-ph/9703442), to appear in ``Heavy Flavours II''
                eds. A.J. Buras, M. Lindner, Advanced Series on Directions in 
                High Energy Physics, World Scientific, Singapore.

\bibitem{B96} A.J. Buras in {\sl Proceedings of the
                  XXVIII Int. Conf. on High Energy Physics} Warsaw, 1996, Z.
                  Ajduk, A.K. Wr\'oblewski eds., World Scientific Singapore
                  and references therein.

\bibitem{BMMP94} A. J. Buras, M. Misiak, M. M{\"u}nz, S. Pokorski
                 {\sl Nucl. Phys.} {\bf B424} (1994) 374.

\bibitem{INLI} T. Inami, C.S. Lim {\sl Prog. Theor. Phys.} {\bf 65} (1981) 
               297, (E.) {\sl ibid.} {\bf 65} (1981) 1772.

\bibitem{BEBOMARI} S. Bertolini, F. Borzumati, A. Masiero,
		   G. Ridolfi {\sl Nucl. Phys.} {\bf B353} (1991) 591;
                   R. Barbieri, G.-F. Giudice {\sl Phys. Lett.} {\bf 309B} 
                   (1993) 86;
                   P. Cho, M. Misiak, D. Wyler, {\sl Phys. Rev.} {\bf D54}
                   (1996) 3329.

\bibitem{AY94} K. Adel, Y.P. Yao {\sl Phys. Rev.} {\bf D49} (1994) 4945.

\bibitem{A94} H. Anlauf {\sl Nucl. Phys.} {\bf B430} (1994) 245.

\bibitem{CMM96} K. Chetyrkin, M. Misiak, M. M{\"u}nz
                {\sl Phys. Lett.} {\bf 400B} (1997) 206.

\bibitem{GHW96} C. Greub, T. Hurth, D. Wyler
                {\sl Phys. Lett.} {\bf 380B} (1996) 385, 
                {\sl Phys. Rev.} {\bf D54} (1996) 3350.

\bibitem{AG95} A. Ali, C. Greub {\sl Phys. Lett.} {\bf 361B} (1995) 146.

\bibitem{FLS94} A.F. Falk, M. Luke, M. Savage {\sl Phys. Rev.} {\bf D49}
                (1994) 3367;
                M. Neubert  {\sl Phys. Rev.} {\bf D49} (1994) 4623.

\bibitem{CLEO95} M.S. Alam et al. {\sl Phys. Rev. Lett.} {\bf 74} (1995) 
                 2885.

\bibitem{MYMH} P.H. Chankowski, S. Pokorski, J. Rosiek {\sl Phys. Lett.} 
               {\bf 274B} (1992) 191; 
               A. Brignole {\sl Phys. Lett.} {\bf 277B} (1992) 313. 



\bibitem{BA} T. Banks {\sl Nucl. Phys.} {\bf B303} (1988) 172;
             M. Olechowski, S. Pokorski {\sl Phys. Lett.} {\bf 214B} 
             (1988) 393;
             S. Pokorski in {\sl Proc. XII Int. Workshop on Weak Interactions 
             and Neutrinos}, Ginosar, Israel {\sl Nucl. Phys. Proc. Suppl.} 
             {\bf B13} (1990) 606;
             S. Dimopoulos, L.J. Hall, S. Raby {\sl Phys. Rev. Lett.} {\bf 68}
             (1992) 1984, {\sl Phys. Rev.} {\bf D45} (1992) 4192;
             G. W. Anderson, S. Raby, S. Dimopoulos, L.J. Hall 
             {\sl Phys. Rev.} {\bf D49} (1994) 3660;
             H.-P. Nilles in Proc. of the 1990 Theoretical Advanced Study 
             Institute in Elementary Particle Physics, eds. M. Cveti\v c, P.
             Langacker, World Scientific, Singapore, p. 633;
             W. Majerotto, B. M\"osslacher {\sl Z. Phys.} {\bf C48} (1990) 
             273;
             M. Drees, M.M. Nojiri {\sl Nucl. Phys.} {\bf B369} (1992) 54;
             B. Ananhtarayan, G. Lazarides, Q. Shafi {\sl Phys. Lett.} 
             {\bf 300B} (1993) 245.

\bibitem{HARASA} L.J. Hall, R. Ratazzi, U. Sarid {\sl Phys. Rev.} {\bf D50}
                 (1994) 7048.

\bibitem{HEMPF} R. Hempfling {\sl Phys. Rev.} {\bf D49} (1994) 6168.

\bibitem{HEMPF_PRIV} R. Hempfling private comunication.

\bibitem{KRPO} P. Krawczyk, S. Pokorski {\sl Phys. Rev. Lett.} {\bf 60} (1988)
               182;
               G. Isidori {\sl Phys. Lett.} {\bf 298B} (1993) 409;
               Y. Grossman, H.E. Haber, Y. Nir {\sl Phys. Lett.} {\bf 357B}
               (1995) 630.

\bibitem{SOLABCTAU} J.A. Coarasa, R.A. Jim\'enez, J. Sol\`a preprint 
                    UAB-FT-407 (hep-ph/9701392). 

\bibitem{SOLAHTB} R.A. Jim\'enez, J. Sol\`a {\sl Phys. Lett.} {\bf 389B}
                  (1996) 53.

\bibitem{SOLATHB} J.A. Coarasa, D. Garcia, J. Guasch, R.A. Jim\'enez, 
                  J. Sol\`a preprint UAB-FT-397 (hep-ph/9607485).

\bibitem{CAOLPOWA} M. Carena, M. Olechowski, S. Pokorski, C.E.M. Wagner
                   {\sl Nucl. Phys.} {\bf B426} (1994) 269;
                   R. Hempfling {\sl Z. Phys.} {\bf C63} (1994) 309;

\bibitem{BOOLPO} F. Borzumati, M. Olechowski, S. Pokorski {\sl Phys. Lett.}
                 {\bf 349B} (1995) 311.

\bibitem{RASA} R. Rattazzi, U. Sarid preprint CERN-TH/96-349 (hep-ph/9612464).

\bibitem{BLPORA} T. Bla\v zek, S. Pokorski, S. Raby {\sl Phys. Rev.} {\bf D52}
                 (1995) 4151.

\bibitem{CMSSM} G.L. Kane, C. Kolda, L. Roszkowski, J.D. Wells {\sl Phys. 
                Rev.} {\bf D49} (1994) 6173, {\bf D50} (1994) 3498;
                M. Olechowski, S. Pokorski {\sl Nucl. Phys.} {\bf B404} (1993)
                590;
                M. Carena, M. Olechowski, S. Pokorski, C.E.M. Wagner  
                {\sl Nucl. Phys.} {\bf B419} (1994) 213;

\bibitem{WAGN} M. Carena, C.E.M. Wagner {\sl Nucl. Phys.} {\bf B452} (1995);
               T. Bla\v zek, M. Carena, S. Raby, C.E.M. Wagner 
               {\sl Nucl. Phys. Proc. Suppl.}  {\bf 52A} (1997) 133;
               J.D. Wells, C. Kolda, G.L. Kane {\sl Phys. Lett.}
               {\bf 338B} (1994) 219;
               D. Pierce, J. Bagger, K. Matchev, R. Zhang  {\sl Nucl. Phys.}
               {\bf B491} (1996) 3;
               H. Baer, M. Brhlik {\sl Phys. Rev.} {\bf D55} (1997) 3201;
               N.G. Deshpande, B. Dutta, S. Oh {\sl Phys. Rev.} {\bf D56}
               (1997) 519.

\end{thebibliography}
\end{document}